\documentstyle[11pt,paspconf]{article}

\begin{document}

\title{Globular Cluster Ages and Str\"omgren CCD Photometry}

\author{Frank Grundahl\altaffilmark{1}}
\affil{University of Victoria, Department of Physics and Astronomy \\
       PO Box 3055, Victoria, BC, V8W 3P6, Canada}

\altaffiltext{1}{Guest worker at the Dominion Astrophysical Observatory, 
Herzberg Institute of Astrophysics, National Research Council of Canada}

\begin{abstract}
Str\"omgren $uvby$ CCD photometry can be used in a variety of ways 
to constrain the absolute and relative ages of globular clusters. 
The reddening corrected $v-y, c_1$ diagram offers the means to 
derive ages that are completely independent of distance. Very 
precise differential ages for clusters of the same chemical 
composition may also be determined from such 2--color plots, or 
from measurements of the magnitude difference, $\Delta u$, between 
the subgiant and horizontal branches on the $u-y, u$ plane (where 
both of these features are flat and well-defined, even for clusters 
like M13 that have extremely blue HBs on the $B-V, V$ diagram). 
Based on high-quality photometry we find that: (1) M92 is $\simeq 
15\,$Gyr old, (2) M3 and M13 differ in age by $< 1\,$Gyr, and (3) 
NGC 288, NGC 362, and NGC 1851 are coeval to within $\sim 1.5\,$Gyr.
These results strongly suggest that age cannot be {\em the} ``second 
parameter''. Finally, we suggest that the observed variations in 
$c_1$ among giant branch stars in all the metal--poor clusters that 
we have studied so far are likely due to star--to--star C and N 
abundance variations, and potentially indicate that most (if not all) 
globular clusters have ``primordial'' variations in at least these 
elements.
\end{abstract}


\section{Introduction}

Str\"omgren photometry has, for many years, served as an important 
source of information about the Galactic field and halo star 
populations due to its capability for deriving such physical 
parameters as metal abundance, effective temperature and luminosity 
for individual stars (see eg. the series of papers by Schuster 
and Nissen (1989), for recent examples and references). The $uvby$ 
filters are of intermediate width and thus their application in 
photoelectric photometry of faint stars, as found near the turnoff 
(TO) in open and globular clusters is rather difficult.  With the 
availability of CCD detectors with excellent near UV sensitivity 
and low readout noise, as well as telescopes with superb image 
quality, it is now possible to obtain precise and accurate 
photometry for individual TO and main--sequence stars in such 
objects. Pioneering work in this field has been carried out by 
Anthony-Twarog (1987a) and Anthony-Twarog \& Twarog (1987b) but 
those early studies were hampered by the low quantum efficiency 
(in particular in the $u$ band centered at 3500\AA) and the high 
readout noise of the available CCD detectors.

In the present contribution some results from a large programme of 
$uvby$ photometry in globular (and open) clusters carried out with 
the Nordic Optical Telescope (NOT) on La Palma and the Danish 1.54m 
telescope at ESO,  will be described.  Other results from this 
programme are described by Landsman et al. (1999, these proceedings) 
and Grundahl et al.  (1998, 1999), who give brief descriptions of the 
data sets that were obtained.

\section{Absolute Cluster Ages}

The motivations for studying star clusters are many and well 
known and they have been discussed at length in the recent literature, 
eg. Renzini \& Fusi Pecci (1988), VandenBerg, Bolte \& Stetson 
(1996). Of particular interest is the determination of absolute 
cluster ages as these provide strict lower limits to the age of 
the Universe. Recent attempts (Carretta, Gratton, Clementini \& 
Fusi Pecci 1999 and references therein) based on the results of 
the ESA space astrometry mission {\sf HIPPARCOS} illustrate how 
the determination of cluster ages from comparison of cluster 
color--magnitude diagrams to stellar evolution models depends 
critically on the quality of the available subdwarf parallaxes 
and cluster reddening and abundance determinations.  As has 
been most clearly illustrated by Schuster and Nissen (1989), $uvby$ 
photometry offers the ability to determine distance independent 
stellar ages for stars near the turnoff through the use of the 
$c_1$ index which, for F and G type stars, is sensitive to surface 
gravity (and hence evolutionary status). Not only is the determination 
of the age independent of distance, but it is also only very slightly 
dependent on reddening because $c_0\,=\,c_1\,-\,0.2\,$E$(b-y)$, and thus 
even relatively large errors in E($b-y$) have only  minor effects 
on $c_1$, and the derived absolute age depends primarily on the 
$c_0$ value of the TO stars.

\begin{figure}
\vspace{3.00in}
\includegraphics{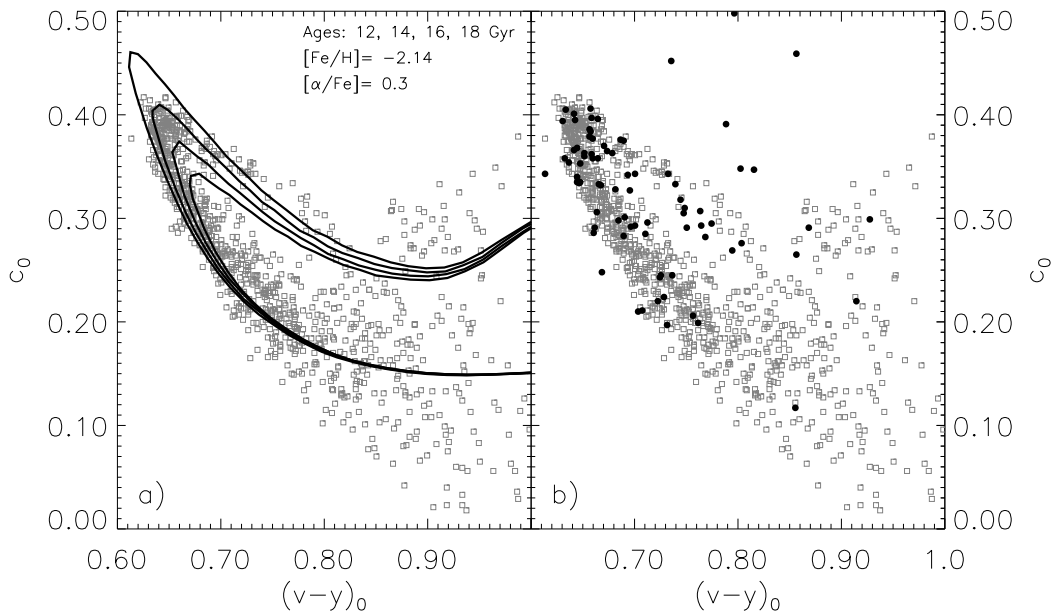}
\caption{a) The reddening corrected $v-y, c_1$ diagram for M92 with 
         isochrones for ages 12, 14, 16, 18 Gyr overplotted. b) Same 
         as a), but with $uvby$ photometry for very metal--poor stars 
         from Schuster et al. (1996) overplotted as filled circles. 
         In both plots E$(b-y)\,=\,0.02$ has been assumed.} 
         \label{fig-1}
\end{figure}

The application of this method is illustrated in Figure 1a for one 
of the most metal--poor GCs, M92, using data obtained with the Nordic 
Optical Telescope in excellent seeing conditions. We note that a 
fairly high age of nearly 15$\,$Gyr seems appropriate for this cluster.
The overplotted isochrones are based on the evolutionary tracks 
reported by VandenBerg et al. (1999), with $uvby$ colors kindly 
calculated by R. A. Bell.  In Fig. 1b  the position of 
very metal--poor field stars from Schuster et al.  (1996) in the 
$v-y, c_1$ diagram is illustrated. We have assumed that the field 
stars have the same reddening as the cluster stars, and not used 
the individual reddenings tabulated by Schuster et al. (1996). This 
is for the reason that for the majority of the stars (most of which 
are found at high galactic lattitudes) have rather large errors in 
E($b-y$) due to relatively large (random) errors in the photoelectric
H$\beta$ photometry of these faint stars. In fact, if the individual 
reddenings are used, the scatter among the field stars increases 
significantly, indicating that the assumption of a constant, low, 
reddening is not unreasonable.

It is worth noting that, as for other methods attempting absolute age
determinations, accurate photometry is essential. However {\em if} this
is available the main uncertainties in the age determinations lie in the
models and their transformation to the observed plane as well as in the 
determination of detailed cluster abundances. Importantly, the problem 
of determining accurate distances is completely eliminated.

\section{Relative Cluster Ages for 2'nd Parameter Pairs}

As we have in hand accurate photometry for M3 and M13, it is 
obviously of interest to compare their ages. These two clusters 
have very similar metallicities but quite different HB morphologies 
with that of M13 being much bluer than M3.  Given that the two 
clusters have a similar [Fe/H]  (the first parameter) what other 
factor(s) govern their HB morphology (the second 
parameter problem)?  Age has been a popular candidate in recent 
years since, according to theoretical calculations older clusters 
have bluer HBs.  However, as discussed by Stetson, VandenBerg \& 
Bolte (1996), there is little solid evidence that this is in fact 
the case --- but at the same time the available photometry for the 
most famous (and easily studied) pairs has not been of a sufficiently 
high quality that unambiguous conclusions could be drawn.  We will 
compare the ages of these two clusters via three different methods:
\smallskip

\noindent
$\bullet$ {\bf{The $\Delta u$--method:}}
It is common to estimate GC ages by deriving the $V$--band luminosity 
difference between the HB and cluster TO ($\Delta V$ method).  However, 
in the case of clusters with different HB morphologies, this is not 
an easy task --- especially if the clusters under study do not have 
stars with an overlap in color on the HB (see Stetson, VandenBerg 
\& Bolte 1996 for a review). We have found that, for BHB clusters, 
the stars hotter than the instability strip and cooler than the 
$u-$jump (Grundahl et al. 1999, Landsman et al., these proceedings) 
the HB is horizontal in a $u-y, u$ diagram, as illustrated in Figure 
2a, where the M3 fiducial sequence has been overlayed on the M13 CMD. 
The shift in color corresponds to $\Delta$E$(B-V)\,=\,$0\fm01.  From 
this figure it is also evident that the SGB is flat, which makes it 
much easier to estimate its level accurately than eg.  $V_{TO}$. 
Thus using $\Delta u^{SGB}_{HB}$ has the potential to yield much 
more accurate estimates of the relative ages of BHB clusters.  Model 
isochrones indicate that $\Delta u^{SGB}_{HB}$ grows at a rate of 
0\fm07$\,$Gyr$^{-1}$.  We see that there appears to be a small 
(0\fm05) difference in $\Delta u^{SGB}_{HB}$ between the two clusters 
corresponding to $0.7\pm0.2\,$Gyr, with M13 being the older (Fig. 
2a). The error estimate is based entirely on the errors of the mean 
values of the HB and SGB levels and relies on the assumption that 
only the age varies between the two. 
\smallskip

\begin{figure}
\vspace{3.00in}
\includegraphics{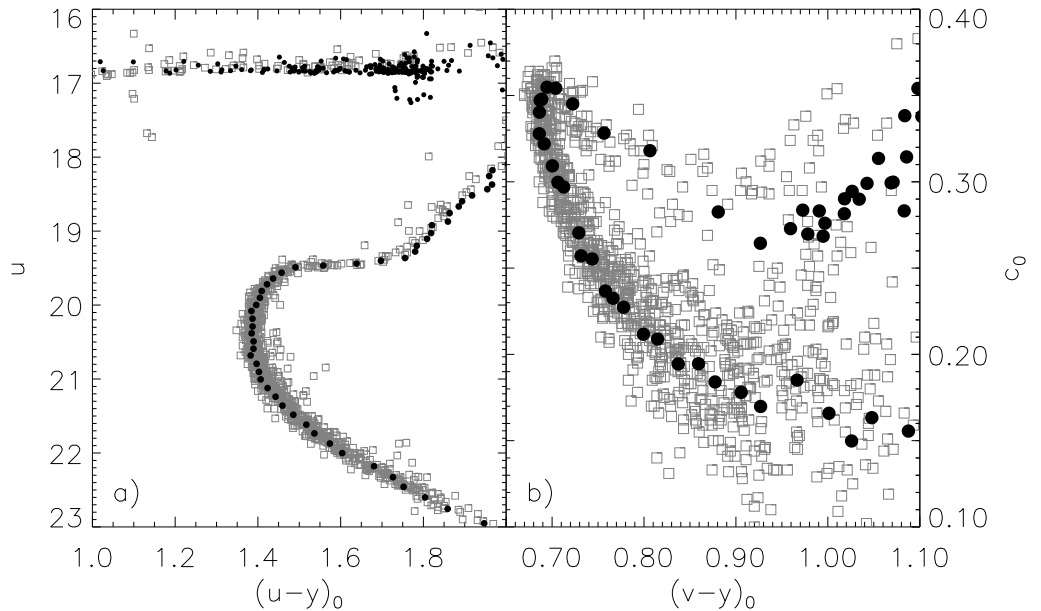}
\caption{Comparison of the CMD's for M3 and M13. In both plots, the
         filled black circles denote the M3 fiducial sequence, and 
         the open gray squares the observed M13 stars. a) The $u-y, 
         u$ diagram.  b) The $v-y, c_1$ diagram, corrected for 
         reddening.} \label{fig-2}
\end{figure}

\noindent
$\bullet$ {{\bf The {\em horizontal} method:}}
VandenBerg, Bolte and Stetson (1990) showed how the color difference
between the cluster turnoff and lower RGB could be used as a relative
age indicator. We have measured the color difference for these two
clusters using their derived fiducial sequences and find that it 
indicates an age difference of $\simeq\,$1~Gyr. This is consistent with 
the $\Delta u^{SGB}_{HB}$ estimate.
\smallskip

\noindent
$\bullet$ {{\bf The $c_1$ value for the cluster turnoff:}} As 
discussed in the previous section, the $c_1$ value for a cluster 
turnoff is an excellent age indicator, independent of cluster distance,
HB-- and RGB morphology, and nearly independent of reddening. In Figure
2b we present a comparison of the fiducials for these two clusters in
the $(v-y)_0, c_0$ plane. It is evident, since the turnoffs coincide, 
that this comparison indicates essentially no age difference between 
the clusters. We note that only shifts based on their canonical 
reddenings have been applied. Furthermore using 
other colors, such as $u-y$, $u-b$ or $b-y$ instead of $v-y$ leads to 
exactly the same conclusion (using only shifts corresponding to a 
reddening difference of $\Delta$E$(B-V) = 0\fm01$) that the age 
difference between these two clusters is essentially zero.  Based on 
the above analyses, which all indicate $\delta(age) < 1\,$Gyr, age 
can hardly be the main factor responsible for the difference in 
the HB morphology of these two clusters.  It should however be 
stressed that this conclusion depends on the assumption that
both clusters have identical abundances.
\smallskip

As part of this programme, we have also observed NGC~288, NGC~362 
and NGC~1851 which are belived to be chemically very similar. An 
accurate estimate of the age differences between them would be 
interesting since the difference in HB morphology between NGC~288 
and NGC~362 is even more extreme than in the case of M3/M13. 
NGC~362 has a much redder HB than NGC~288, whereas the morphology 
of the HB in NGC~1851 is intermediate between the two.  Unfortunately 
the $u$ photometry for these clusters is not precise  enough to use 
$c_1$ to estimate their age differences. However if we use the 
$u-y, u$ diagram and NGC~1851 as a ``bridge'' between the HB of 
NGC~288 and NGC~362, as first proposed by Stetson, VandenBerg
and Bolte (1996), we can estimate their relative ages. This
approach is shown in Figure 3. 

\begin{figure}
\vspace{3.00in}
\includegraphics{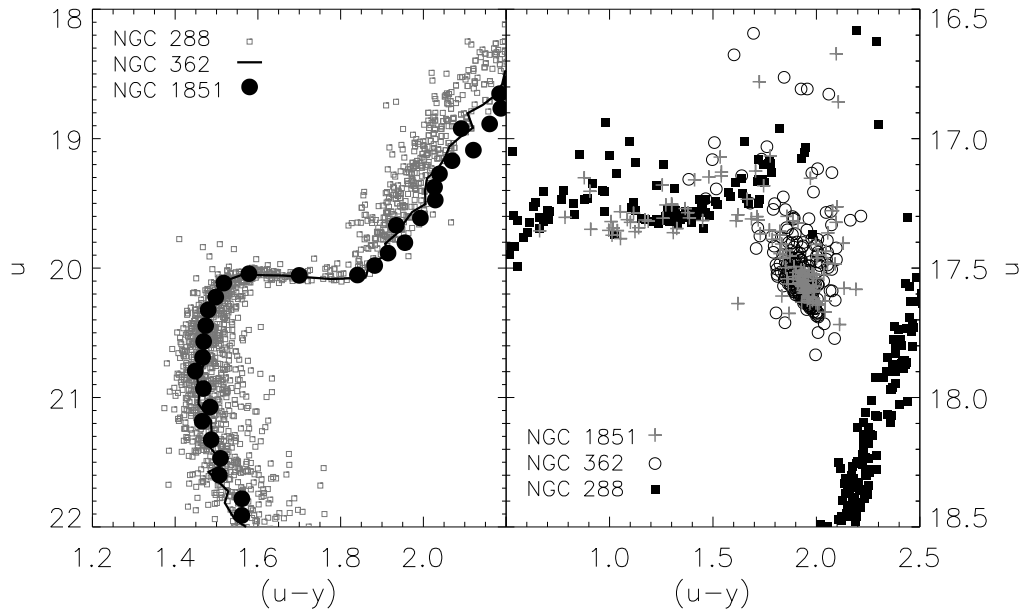}
\caption{A composite diagram showing a comparison between NGC~288, 
         NGC~362 and NGC~1851. Left: The $u-y, u$ diagram. Right: 
         A blow-up of the HB region in the $u-y, u$ plane. Individual 
         HB stars for each of the three clusters have been plotted. 
         Exactly the same offsets as in the left figure has been
         applied to the photometry shown in the right figure.} \label{fig-3}
\end{figure}

We proceeded to make this comparison, by first deriving fiducial 
sequences for the MS, TO and lower RGB of each cluster. Next we added 
arbitrary shifts in color and luminosity to align the TO regions
(Fig. 3, left). 
These shifts were then applied to the photometry for the HB stars in 
each cluster, resulting in the right panel of Figure 3.

From the right panel in Fig. 3 we note that the NGC~1851 stars (gray 
plus signs) match well the ZAHB level for NGC~288 (black squares) in 
the color interval 1--1.5 in $u-y$. Also, at $(u-y)\,\simeq\,1.9$ 
the NGC~1851 RHB stars match the NGC~362 (open circles) RHB stars; 
the small difference in level (or color) corresponds to 0\fm04 in 
$u$.  Since the offsets applied to the HB stars have been determined 
from the requirement that the age--sensitive SGB regions (left panel) 
are aligned, we conclude from this comparison that the age 
differences between these three clusters are $<\,1.5\,$Gyr. Such a 
small difference is not enough to explain the difference in HB 
morphology between NGC~288 and NGC~362 (Catelan \& De Freitas 
Pacheco 1993).

It is however worth noting that, as was the case for the M3/M13 
comparison, the color difference between the TO and lower RGB seems
to be larger for NGC~362 and NGC~1851 than for NGC~288. As no adequate 
models on the $uvby$ system , at this metallicity, are available at 
this time we cannot determine a limit to the possible age difference 
between the clusters from this method. VandenBerg (1999) discusses 
the possible dependence of the interpretation of age differences 
derived via the ``horizontal'' method on abundance and choice of 
filters. 

It seems very likely that the age differences between these three 
clusters are too small to explain the variations in HB morphology 
in terms of age. However, additional observations will be required 
to settle this issue. In particular, we suggest that, if sufficiently 
accurate $c_1$ values can be measured for the cluster TO's, the age 
differences can be inferred independently of the HB morphology (as 
for M3/M13). Such a project, which  is easily carried out in 1--2 
nights on a 4m class telescope avoids completely the difficult 
problem of comparing a red HB to a blue HB.

\section{RGB $c_1$ scatter}

Grundahl et al.~(1998) reported a large spread in $c_1$ (at fixed 
luminosity or color) among the RGB stars in M13 and speculated 
that this could be due to star-to-star variations in the strength 
of their CN bands.  We have subsequently analysed $uvby$ photometry 
for many other clusters and in all cases we find a $c_1$ scatter 
reminicent to that found in M13. For clusters in the metallicity 
range $-1.2$ to $-1.7$ we also find an indication that the ``width'' 
of the $c_1$ scatter gets smaller as the luminosity increases from 
the RGB bump level towards the RGB tip.

\begin{figure}
\vspace{3.00in}
\includegraphics{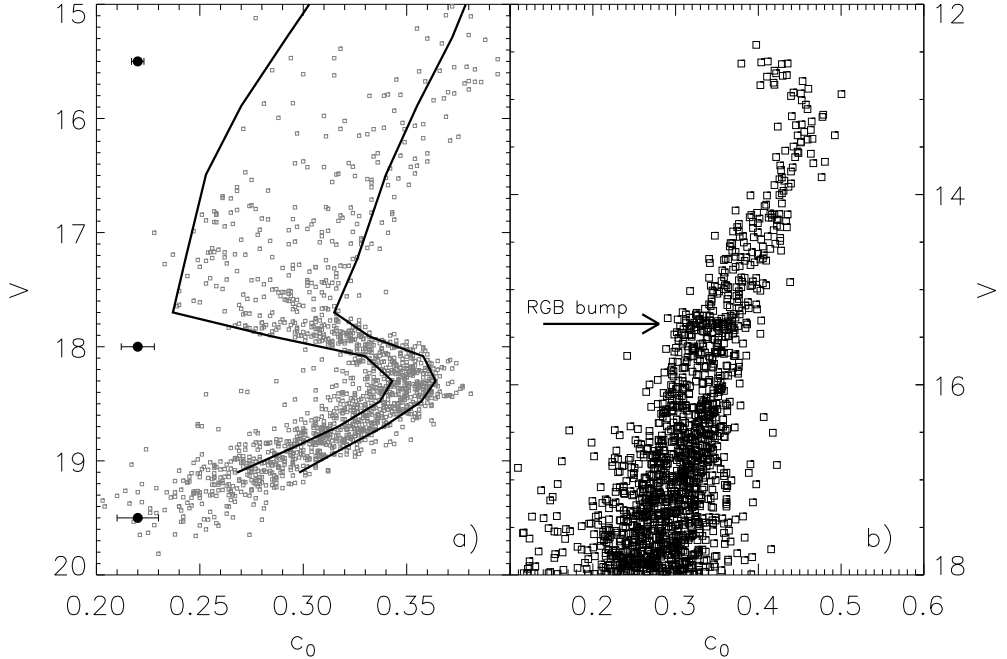}
\caption{Illustration of the observed $c_1$ scatter in M13 and M3.
         a) M13 with models (full drawn lines) assuming a spread 
         in C and N similar to that observed in MSTO stars in NGC 
         6752 and 47~Tuc (from M. Briley, private communication) 
         overplotted. b) The observed $c_1$ spread in M3, illustrating 
         how it becomes smaller for stars brighter than the RGB bump.}
\end{figure}

Michael Briley has kindly carried out simulations of RGB spectra 
with varying amounts of C and N abundances corresponding to that 
observed in TO and MS stars in 47~Tuc and NGC~6752. We show in 
Figure 4a how these simulations reproduce the width of the $c_1$ 
scatter observed in M13. The variations found in these two clusters 
are widely suspected to be ``primordial'' in origin since there 
are no known mixing proceses that can operate in stars near the 
MSTO in GC's which could produce star--to--star variations in C 
and N sufficiently large to produce the observed $c_1$ scatter.  
If C and N variations are the explanation for the observed scatter,
then this probably implies that most GC's have significant 
variations in (at least) C and N, due to earlier generations of 
field- and/or cluster stars. We emphasize that this interpretation 
relies on the correctness of the atmosphere models as we do not 
yet have spectra for the stars.  The observed ``narrowing'' of 
the $c_1$ scatter in stars brighter than the RGB bump (see Fig. 
4b) could potentially be understood as the result of deep mixing 
commencing at this point on the RGB (Sweigart \& Mengel 1979, 
Charbonnel, Brown \& Wallerstein 1998, Carretta, Gratton, Sneden 
\& Bragaglia 1998).  Such mixing would bring C$\,\rightarrow\,$N 
processed material to the surface of RGB stars. This would have 
the effect of decreasing the C abundance and increasing the N 
abundance with the net result that the CN and CH features present 
in the $v$ and $u$ bands become weaker, thereby reducing the 
observed $c_1$ spread.

\section{Conclusions}

Str\"omgren photometry offers the potential to determine globular
cluster 
ages without needing to know their distances and with only a weak 
sensitivity on the cluster reddening. If current steller interior 
and atmosphere models are correct, this implies an age of nearly 15
billion years for M92. For two pairs of second parameter clusters 
we showed how the inclusion of the $u$ filter can help in measuring 
their relative ages, and found that M3 and M13 differ by less than 
one billion years in age. The range in age encompassed by  NGC~288, 
NGC~362 and NGC~1851 appears to be less than 1.5 billion years. To 
improve further on these age constraints requires a better knowledge 
of the detailed cluster abundances.  Finally, it was discussed that 
the $c_1$ index is potentially an excellent indicator of the surface 
abundances of C and N in low metallicity RGB stars -- and the results 
for clusters that we have observed so far indicate that they may all 
have ``primordial'' variations in (at least) these elements and that 
the reduction of the $c_1$ scatter observed in RGB stars brighter 
than the bump could be due to non--canonical mixing.

\acknowledgments

I would like to acknowledge useful and stimulating conversations 
and exchange of data and models with Michael Andersen, Roger Bell, 
Michael Briley, M\'arcio Catelan, James Hesser, Poul Erik Nissen, 
Peter Stetson and Don VandenBerg. Financial support from The Danish 
Natural Sciences Research Council, Don VandenBerg, The Herzberg Institute 
of Astrophysics and the Carlsberg Foundation is also gratefully
acknowledged. 

% ===========================================================================


\begin{references}

\reference Anthony--Twarog, B. J. 1987a, \aj, 93, 1454
\reference Anthony--Twarog, B. J., \& Twarog, B. A. 1987b, \aj, 94, 1222
\reference Carretta, E., Gratton, R., Sneden, C., \& Bragaglia, A., to 
           appear in the proceedings of the meeting ``Galaxy Evolution:
           Connecting the Distant Universe with the Local Fossil 
           Record'', Observatoire de Paris--Meudon, 21--25 September 1998
\reference Carretta, E., Gratton, R., Clementini, G. \& Fusi Pecci, F., 
           1999, preprint, (astro-ph 9902086)
\reference Catelan, M., \& De Freitas Pacheco, J. M., 1993, \aj, 106, 1858
\reference Charbonnel, C., Brown, J. A., Wallerstein, G., 1998, A\&A, 332, 204
\reference Grundahl, F., VandenBerg, D. A., \& Andersen, M. I., 1998, 
           \apjlett, 500, L179
\reference Grundahl, F., Catelan, M., Landsman, W. B., Stetson, P. B., 
           \& Andersen, M. I., 1999, ApJ, 524, Oct. 10 issue. 
\reference Landsman, W. B. 1999, these proceedings.
\reference Renzini, A., \& Fusi Pecci, F., 1988, ARA\&A, 26, 199
\reference Schuster, W. J., \& Nissen, P. E., 1989, A\&A, 221, 65
\reference Schuster, W. J., \& Nissen, P. E., Parraro, L, Beers, T. C., 
           \& Overgaard, L. P., 1996, A\&AS, 117, 317
\reference Stetson, P. B., VandenBerg, D. A., \& Bolte, M., 1996, \pasp, 
           108, 560
\reference Sweigart, A. V., \& Mengel, J. G., 1979, \apj, 229, 624
\reference VandenBerg, D. A., 1999, ApJ, submitted
\reference VandenBerg, D. A., Bolte, M., \& Stetson, P. B., 1990, \aj, 
           100, 445
\reference VandenBerg, D. A., Bolte, M., \& Stetson, P. B., 1996, ARA\&A, 
           34, 461
\reference VandenBerg, D. A., Swenson, F. J., Rogers, J. J., Iglesias, C. A., 
           \& Alexander, D. R., 1999, \apj, submitted

\end{references}
\end{document}